\def\sss{\scriptscriptstyle}
\def\beq{\begin{equation}}
\def\eeq{\end{equation}}
\def\beqa{\begin{eqnarray}}
\def\eeqa{\end{eqnarray}}
\def\darr{\mathrel{\hbox{\rlap{\hbox{\lower1.0pt\hbox{$\partial_\mu$}}}\hbox{\raise5.0pt\hbox{$\!\leftrightarrow$}}}\!}}
\def\figeight{\bigcirc\!\bigcirc}
\def\ptp#1#2#3{{Prog.\ Theor.\ Phys.\ }{{\bf #1} {(#2)} {#3}}}
\def\setsun{{\bigcirc\phantom{|}\!\!\!\!\!\!\!\!\!-\!\!\!\!-}}
\def\avg#1{{\left\langle #1 \right\rangle}}
\def\lsim{\mbox{\raisebox{-.6ex}{~$\stackrel{<}{\sim}$~}}}

\documentstyle[pramana,epsf,floats]{ias}
\begin{document}
\mark{{Status of Electroweak Phase 
Transition and Baryogenesis}{J.M.\ Cline}}
\title{Status of Electroweak Phase 
Transition and\\ Baryogenesis}

\author{J.M.\ Cline}
\address{McGill University, Dept.\ of Physics, 3600 University
St., Montre\'eal, Qu\'ebec H3A 2T8, Canada}
\pacs{98.80.Cq \hfill McGill/00-09}
\abstract{I review recent progress on the electroweak phase transition
and baryogenesis, focusing on the minimal supersymmetric standard
model as the source of new physics.}

\maketitle
\section{Cosmological Phase Transitions--Electroweak}

It is possible that the universe has undergone a number of phase
transitions, as illustrated in Table 1.  In most cases, it is difficult
to find a signature of such a transition which survives to the present
day.  One important class of exceptions is when (meta)stable
topological defects like cosmic stringes are formed; I will not deal
with this important topic in the present talk.  Instead I will focus on
the other main possibility of interest, the case of a first order
transition.

\begin{table}
\begin{tabular}{||c|c|c||}
Energy Scale & Transition & Order Parameter \\ \hline \hline
$M_{\rm Planck}$ & spacetime foam  $\to$ classical spacetime
& $\langle g_{\mu\nu} \rangle$ \\
\hline
$M_{\rm \sss GUT}$ & GUT symmetry $\to$ standard model & $\langle
\hbox{GUT Higgs}\rangle$ \\ 
\hline
Intermediate scale & Inflation & $\langle \hbox{Inflaton}\rangle$\\
\hline
$M_{\rm \sss SUSY}$ & SUSY Breaking & 
$\langle (\hbox{gaugino})^2\rangle$? \\
\hline
100 GeV & Electroweak: SU(2)$\times$U(1)$\to$U(1)$_{\rm em}$ &
$\langle \hbox{SM Higgs}\rangle$ \\
\hline
100 MeV & QCD chiral phase transition & $\langle \bar q_L q_R \rangle$\\
\end{tabular}
\caption{Some phase transitions which may have taken place
during cosmological history.}
\end{table}

A memorable example of the effect a first order phase transition could
have was proposed by Witten [1] in 1984.  He noted that when
bubbles of the chiral symmetry broken phase form (where $\langle \bar q_L
q_R \rangle \neq 0$),  baryons tend to pile up on the bubble walls.
Neutrons diffuse quickly into the bubble interiors, but protons diffuse
more slowly, and the spatial separation of isospin was found to have an
observable effect on helium production in primordial nucleosynthesis.
Unfortunately, lattice studies have shown that the QCD chiral phase
transition is a smooth crossover when quarks have masses, so that in
fact there is no bubble formation.

If the QCD transition was not first order, what about the next lowest
energy example on our list, the electroweak phase transition (EWPT)?
In this case the bubbles would contain regions of nonvanishing Higgs
field VEV, $\langle H \rangle \neq 0$.  If CP is violated on the bubble
wall, there can be a pile-up of chiral charge, which biases the
anomalous sphaleron interactions of the standard model to produce
baryons.  We could thus explain the baryon asymmetry of the universe.
Unfortunately, the lattice gauge theorists have again spoiled our fun
by finding that for Higgs masses $m_H > 70-80$ GeV, the transition is a
smooth crossover.  (The latest limit from LEP is $m_H > 106$ GeV.)
However, it is much easier to change this negative conclusion by adding
new physics (like supersymmetry) to the electroweak theory than it is
for QCD.  This will be the subject of the rest of this talk.

\section{How strong is the EWPT?}

One of the main tools for studying the strength of the EWPT analytically
is the finite temperature effective potential, defined by
\beq
e^{-\beta \int d^3x\, V_{\rm eff}[\Phi]} = 
\int \prod_i {\cal D}\phi_i \, e^{-\int_0^\beta d\tau \int d^3\!x\,
	 S[\Phi,\phi_i]/\hbar}
\eeq
It is a path integral over fluctuating fields $\phi_i$, around a
constant background field $\Phi$, in our case the Higgs field.
The fields are in imaginary time with periodic boundary conditions
(for bosons; antiperiodic for fermions) between $\tau=0$ and $\tau=
\beta=1/T$.  $V_{\rm eff}$ can be computed in perturbation theory,
represented by Feynman diagrams like $\bigcirc$ at one loop, and 
$\figeight$ or $\setsun$ at two loops.  The one loop term is the
effect of a noninteracting boson or fermion gas, of which the 
particle masses depend on the background field $\Phi$,
\beq
	V_{\rm 1-loop} = T\sum_i\mp\int{d^3\!p\over (2\pi)^3}
	\ln\left(1\pm e^{-\sqrt{p^2+m^2_i(\Phi)}/T}\right),
\eeq
where the upper (lower) sign is for fermions (bosons) in the loop.
This can be approximated in a high-temperature expansion as
\beqa
&&\sum_i {m^2_i(\Phi)T^2\over 48} \hbox{($\times 2$ for bosons)} \sim \Phi^2 T^2
\nonumber\\
&&- \sum_i {m^3_i(\Phi)T\over 12\pi} \hbox{(bosons only)} \sim -\Phi^3 T
\nonumber\\
&&\mp {m^4_i(\Phi)\over 64\pi^2}\left(\ln{T^2\over\mu^2} + C_i\right)
+ O(m_i^6/T^2)
\eeqa
The $\Phi^2 T^2$ term is responsible for symmetry restoration at high
$T$, while the $-\Phi^3 T$ term gives a barrier or bump
({\epsfysize=0.125in\epsfbox{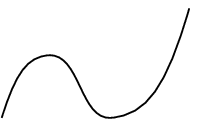}\,) in the potential, which is
responsible for the first order transition, if it occurs.

Unfortunately, perturbation theory can be unreliable, especially near
the phase transition [2].  If one starts with an arbitrarily 
complicated diagram contributing to $V_{n\rm-loop}$ and adds an extra
$W$ boson propagator, the ``cost'' of the new loop is a multiplicative factor
which parametrically has the form
\beq
	\epsilon = g^2 T \int {d^3\!p\over (2\pi)^3}
	\left( p^2 + m^2_W(\Phi)\right)^{-2} \sim g^2 {T\over m_W(\Phi)}
	\sim g{T\over\Phi}.
\eeq
The relevant value of $\Phi$ is $\Phi_c$, the VEV in the broken phase
at the critical temperature where $V_{\rm eff}(0) = V_{\rm eff}(\Phi_c)$.
If we parametrize $V_{\rm eff} 
\cong AT^2\Phi^2 - BT\Phi^3 + \lambda\Phi^4$
then $\Phi_c = 2BT/\lambda \sim g^3T/\lambda \sim (m^2_w/m^2_H)gT$.
Therefore the perturbative parameter $\epsilon$ goes like $\lambda/g^2
\sim m^2_w/m^2_H$, and perturbation theory breaks down for heavy Higgs
bosons, $m_H > m_W$.

There are several ways to combat the breakdown of pertubation theory at
finite temperature.  The most brute force method is to use lattice
gauge simulations for the full 4-D theory [3].  Somewhat
easier is to use dimensional reduction---integrating out the heavy
Matsubara modes (the Fourier modes of the compactified imaginary time
direction, with masses $m_n \sim n\pi T$) to get an effective 3-D
theory [4,5].  The 3-D theory can then be studied on the
lattice [6], much more easily than the 4-D theory.  A third
method is to compute $V_{\rm eff}$ to higher order in perurbation
theory [7--10].  This sounds unjustified, since
pertubation theory was supposed to be breaking down, but experience
shows that in fact it works rather well in the cases of
interest---where the transition is strongly first order.

As mentioned above, the lattice studies have established that, although
there is a line of first order phase transitions in the $T-m_H$ plane
at small $m_H$, it comes to an end (at a point where the transition is
2nd order) around $m_H = 75$ GeV.  For larger $m_H$ there is no clear
distinction between the unbroken and broken phases of the electroweak
theory.  For example, massive $W$ bosons in the broken phase cannot
be distinguished from massive composite objects $(H^\dagger\darr H)$
in the symmetric, confining phase.

To get a first order transition for realistic Higgs masses, we need to
add new physics which couples significantly to the Higgs boson.  Let's
see how supersymmetry can do this.

\section{Adding Supersymmetry}

Recall that at one loop, it is the cubic term, $-T m^3(H)/12\pi$ that
gives bump in the potential ({\epsfysize=0.125in\epsfbox{bump.eps}\,)
 hence a first order transition.  In the
minimal supersymmetric standard model (MSSM), we have two Higgs
doublets, and the top squarks ($\tilde t_L$ and $\tilde t_R$) couple
strongly to the second one, $H_2$.  Ignoring small terms involving
the weak gauge coupling $g$, the $\tilde t_L$-$\tilde t_R$
mass matrix has the form
\beq
	M^2_{\tilde t_L,\tilde t_R} = 
	\left(\begin{array}{cc} m^2_Q  + y^2 H_2^2 & y(A_tH_2-\mu H_1) \\
				y(A_tH_2-\mu H_1)  & m^2_U  + y^2 H_2^2
	\end{array}\right)
\eeq
We need at least one of the SUSY breaking soft masses, $m^2_U$ or
$m^2_Q$, to be small so that there will be an eigenvalue $m(H)$ which
is really cubic, $\sim H^3$, not $(m_0^2 + y^2 H^2)^{3/2}$, since the
latter form does not give a true bump in the potential [8].  On
the other hand, one of $m^2_U$ or $m^2_Q$ should be large so that the
stop radiative correction to the Higgs mass can be big enough to satisfy
the experimental constraint:
\beq
	m^2_H \sim m^2_Z + O\left[ {m_t^4\over v^2}\ln\left(m_{\tilde t_L}
	m_{\tilde t_R}\over m_t^2\right)\right] > (96 \hbox{\ GeV})^2
\eeq
(The limit on $m_H$ in the MSSM is about 10 GeV weaker than in the SM.)
The precision electroweak $\rho$ parameter (a.k.a.\ $\epsilon_1$) dictates 
that the $\tilde t_L$-like squark should be the heavy one, hence the 
$\tilde t_R$-like squark is light.

It turns out that two-loop effects are crucial for getting a strong enough
phase transition[7--10].  Diagrams like $\setsun$, with a gluon or
Higgs boson as one of the internal lines, contribute a term
\beqa
	\Delta V_{\rm eff} &=& -(8g_s^2 - 3y^2\sin^2\beta)T^2 m^2_{\tilde t_R}(H)
	\ln\left({m_{\tilde t_R}(H)\over T}\right)\nonumber\\
 	&\sim & 
	- C T^2H^2\ln\left({H\over T}\right)
\eeqa
to $V_{\rm eff}$, 
whose form is unlike any generated at one loop.  One can show that such a term
shifts the critical value of the Higgs VEV according to
\beq
	{\langle H\rangle\over T} \sim {B\over\lambda} +
	\sqrt{\left({B\over\lambda}\right)^2 + {2C\over \lambda}},
\eeq
where $B$ is the part coming from the cubic term.  $\langle H\rangle/T$ is
the relevant measure of the strength of the transition, as we will discuss
below.

In ref.\ [10] we have performed a Monte Carlo search of the MSSM parameter
space to find those which give a strong enough phase transition for electroweak
baryogenesis.  We can summarize the resulting constraints on the squark
masses as follows:
\beqa
\label{eq:Rlimit}
&&	120 {\rm\ GeV} \lsim m_{\tilde t_R} \lsim m_{\rm top} \\
\label{eq:Llimit}
&&	m_{\tilde t_L} > 265 {\rm\ GeV\ }\times e^{(m_H-95{\rm\ GeV})/9.2}
\eeqa
It is often said that electroweak baryogenesis puts an upper limit on the
Higgs boson mass, but this is not correct in the MSSM, where the light right
stop is doing most of the job of making the transition first order.  However,
as we can see from eq.\ (\ref{eq:Llimit}), it is true that the left stop
quickly becomes unnaturally heavy as $m_H$ is increased.  Thus we are pushed
to a rather strange corner of parameter space, where $\tilde t_R$ is 
extremely light, and $\tilde t_L$ extremely heavy.  If $m_H$ turns out to
be much heavier than its current experimental limit, there must be some other
new physics accounting for its mass.

More promising is the prediction (\ref{eq:Rlimit}) for the stop mass.
Although Tevatron searches for the decay $\tilde t \to \tilde\chi^0_1 c$
are only beginning to probe the region of interest
(fig.\ \ref{fig1}(a)) [11], in Run II this decay will probe up to
the top mass if $m_{\tilde\chi^0_1}$ is in the right range (dark region
of fig.\ \ref{fig1}(b)), and the decays $\tilde t\to b\tilde\chi^+_1$
or $\tilde t\to bW\tilde\chi^0_1$ may be more revealing, if the
chargino is light (lighter regions) [12].

\begin{figure}
\centerline{\epsfxsize=2.75in\epsfbox{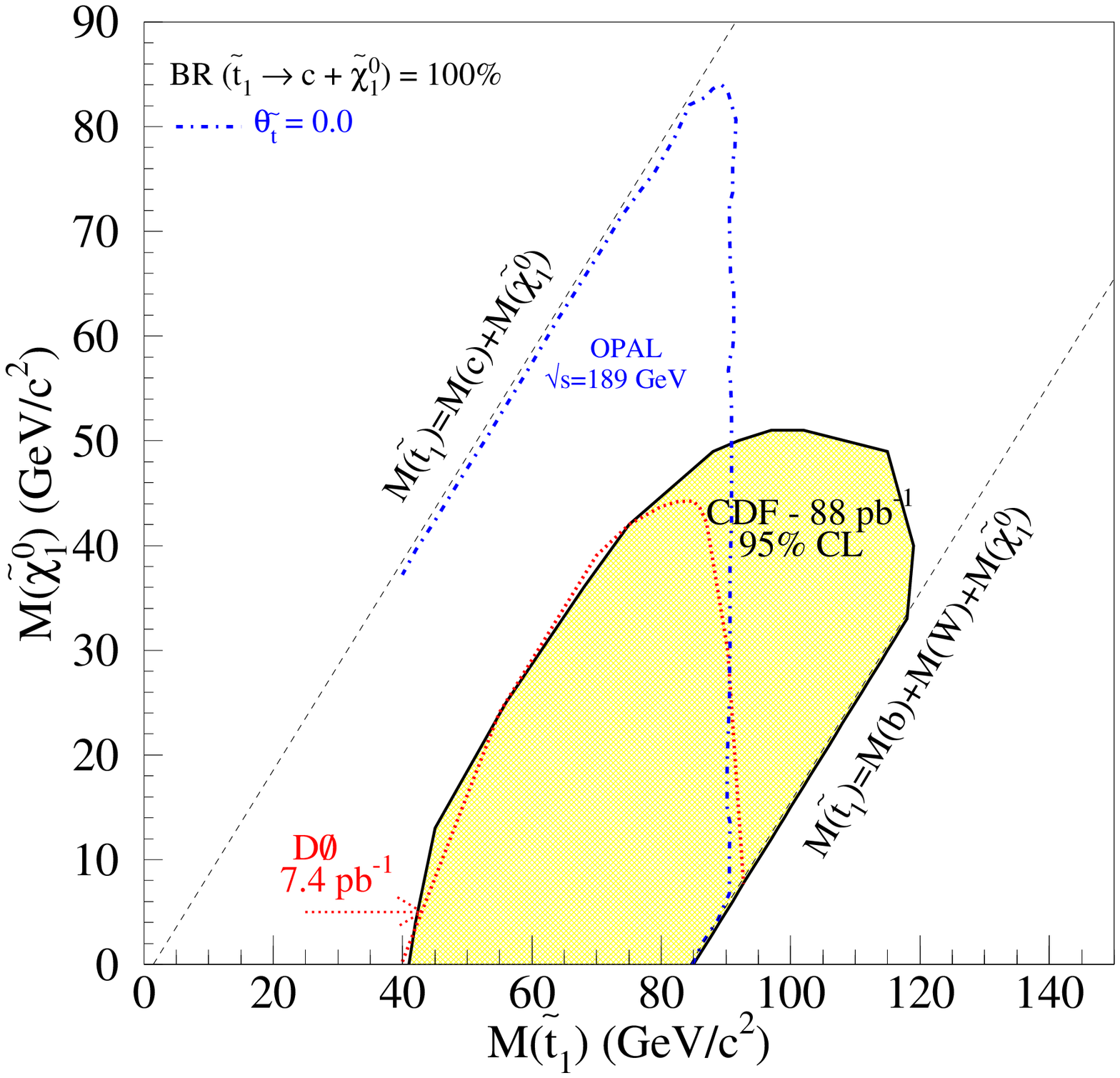}\hfill
\epsfxsize=2.4in\epsfbox{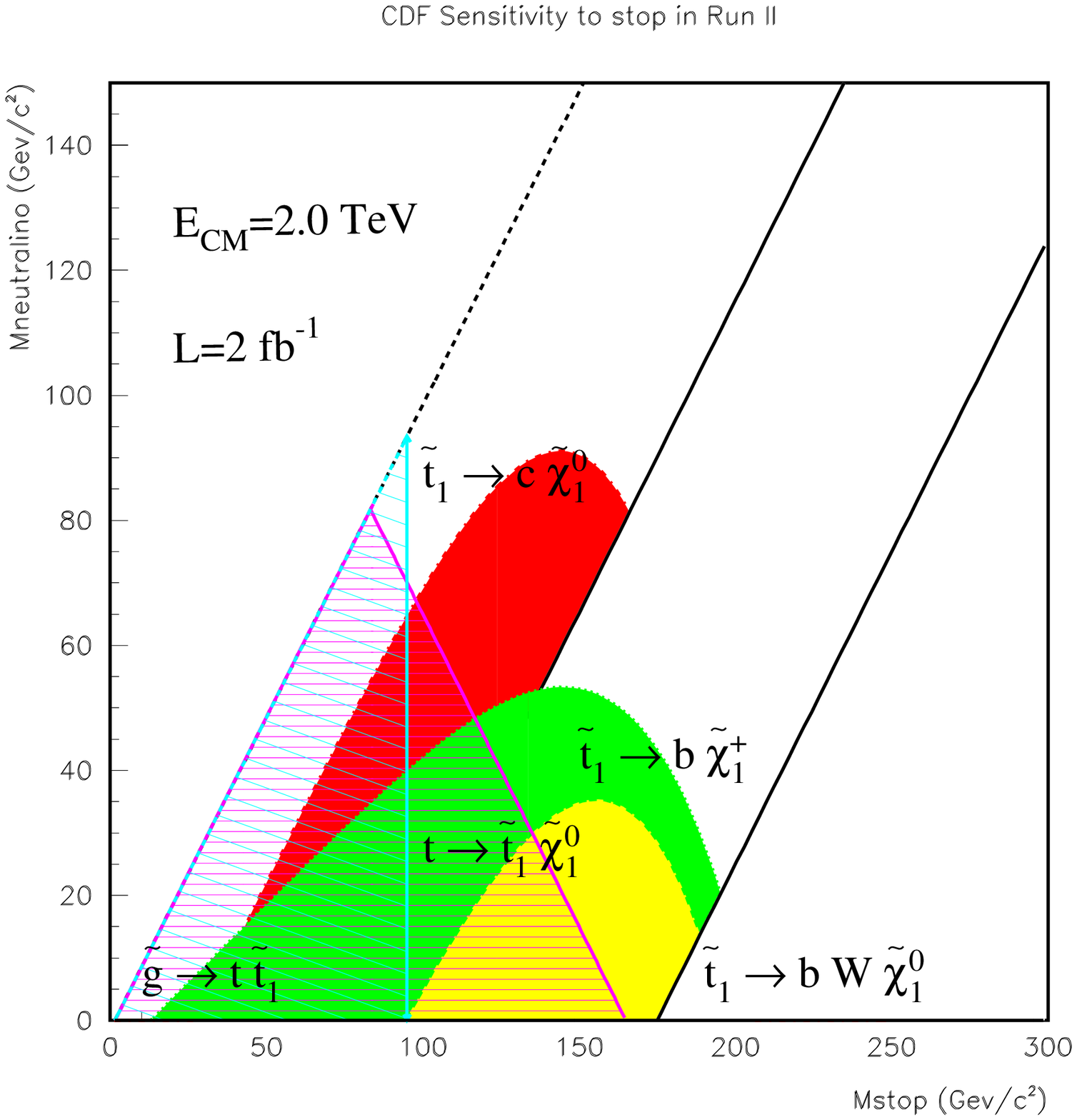}}
\vspace{0.2in}
\caption{(a) Present Tevatron limits [11] 
and (b) sensitivity in Run II for the stop mass [12] 
in the (neutralino mass)--(stop mass) plane.}
\label{fig1}
\end{figure}

An interesting consequence of such a light stop is that its bare
(mass)$^2$ must be negative.  In the absence of left-right mixing,
$m^2_{\tilde t_R} = m^2_U + m^2_t$, so $m_{\tilde t_R} < m_t$ implies
$m^2_U < 0$.  This can cause an instability toward condensation of the
$\tilde t_R$ field in the early universe, when $\langle H\rangle$ is
still zero, which would break SU(3)$_{\rm color}$ [9].  Indeed,
the lattice study of ref.\ [13] finds a phase diagram similar to
fig.\ \ref{fig2}(a), which shows that color-breaking occurs when
$(-m^2_U)^{1/2}$ exceeds $60-70$ GeV, depending on the critical
temperature $T_c$.  In ref.\ [14] we have constructed the two
loop effective potential $V_{\rm eff}(\tilde t_R, H)$ for stop and
Higgs fields, and studied the possibility that the universe might
temporarily enter the color-breaking phase, before finally tunneling to
the EW-breaking vacuum which we inhabit now.  We find that because of
the potential barrier separating the two minima, shown in
fig.\ \ref{fig2}(b), the rate of tunneling is so small that if the
universe ever enters the color-breaking minimum, it stays there
forever.  This conclusion could however change in the presence of
R-parity violating interactions like $y_{332}A\tilde t_R^a \tilde b_R^b
\tilde s_R^c \epsilon_{abc}$ which would lower the barrier.

\begin{figure}
\centerline{\epsfxsize=2.75in\epsfbox{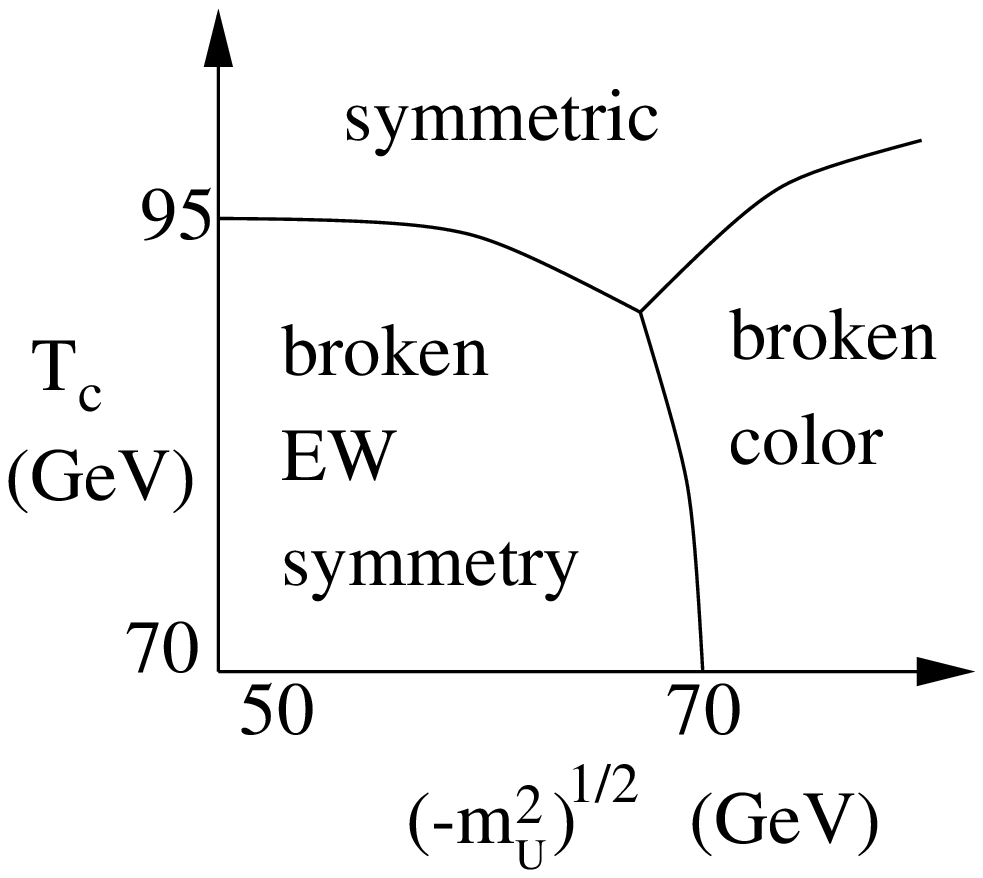}\hfill
\epsfxsize=2.4in\epsfbox{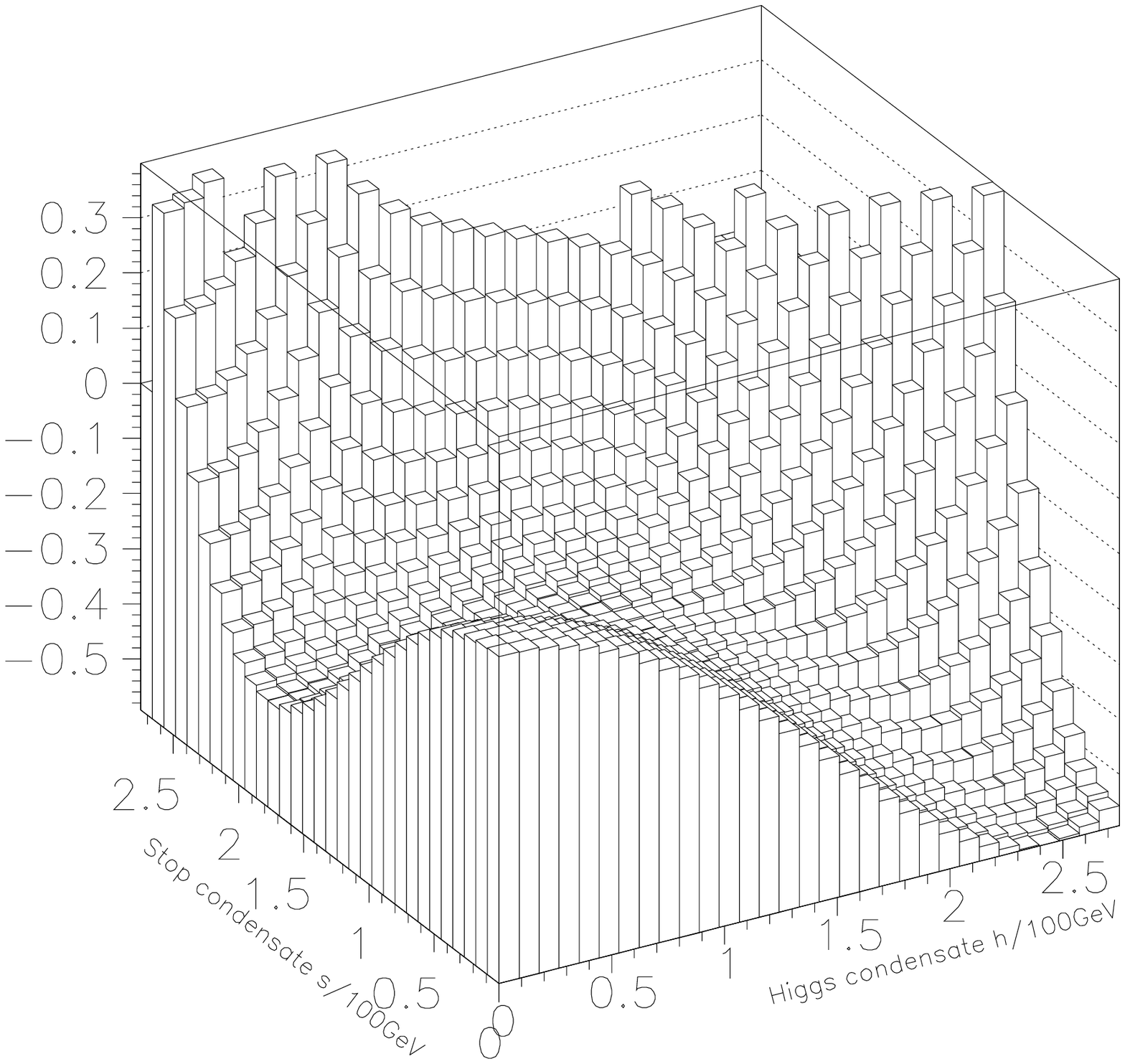}}
\vspace{0.2in}
\caption{(a) Phase diagram
and (b) effective potential, $V_{\rm eff}(\tilde t_R, H)$, 
for color breaking due to stop condensation.}
\label{fig2}
\end{figure}

\section{Baryon Asymmetry of the Universe and Baryogenesis} 

I will now review some basics about baryogenesis and describe some
recent developments, before giving the latest details on electroweak
baryogenesis in the MSSM.  It is very unlikely that we live in a
baryon-antibaryon symmetric universe since there is no evidence of
antigalaxies colliding with galaxies, which would give an intense
source of gamma rays.  If there are  regions consisting of antimatter
outside of our Hubble volume, it is difficult to imagine how separation
from ordinary matter could occur on such a large distance scale.  Big
Bang nucleosynthesis tells us that 
\beq
	2 \quad \lsim \quad\eta_{10} \equiv 
	{n_{\sss B}-n_{\bar{\sss B}}\over n_\gamma}\times 10^{-10}
	\quad \lsim \quad 3
\eeq
 Yet in the early
universe the most natural initial condition is equal numbers of baryons
and antibaryons, since there are many possible $B$-violating
interactions that could be in thermal equilibrium at high temperatures,
for example a dimension 9 operator like $(udd)^2/\Lambda^5$, which
would cause neutron-antineutron oscillations, or the interactions with
heavy $X$ gauge bosons in grand unified theories.  In fact we need not
look so far afield, since sphalerons violate baryon number within the
standard model itself!

It is well known that Sakharov's three conditions must be fulfilled to
generate a baryon asymmetry: (1) baryon number violation; (2) $C$ and
$CP$ violation; (3) loss of thermal equilibrium for the B-violating
interactions.  The first and second conditions are present within the
SM, but (2) is too weak for baryogenesis, and (3) is not fulfilled at
all.  We have already seen how the MSSM can increase $\langle
H\rangle/T$; this is what is needed to make the sphalerons go out of
equilibrium inside the bubbles that form during the EWPT.  The MSSM can
also cure the problem of getting strong enough CP violation, as we will
discuss in the next section.  Here, I will only mention some of the other
proposals for baryogenesis which are new or currently popular.

Baryogenesis via leptogenesis [15] is one of the most plausible
alternatives to electroweak baryogenesis.  In analogy to GUT
baryogenesis, heavy sterile neutrinos decay out of equilibrium,
producing a lepton asymmetry, which is converted by sphalerons into the
baryon asymmetry.  The predictions can be related to neutrino masses in
a GUT framework like SO(10) [16].

The Affleck-Dine mechanism [17] has long been one of the most efficient
baryogenesis mechanisms.  It uses the fact that SUSY scalar potentials from
D-terms can often have flat directions which generate a huge baryon number
when the flatness is lifted and the field evolves by spiraling in the complex
plane.  It was recently pointed out that this can be combined with leptogenesis
to give a minimal supersymmetric baryogenesis model in which the flat direction
is a linear combination of $H_2$ (the second Higgs doublet) and $L_e$ (the
selectron), using physics which is already needed for generating neutrino masses
[18].

Large extra dimensions can present a serious challenge to baryogenesis by
constraining the reheat temperature after inflation to be very low
[19].  The Randall-Sundrum alternative of warped compactification
[20] evades this problem; so perhaps do intrinsically ``braney''
approaches to baryogenesis [21].

Other novel ideas make use of the phase transition in left-right symmetric
models [22] and decaying primordial black holes [23].

\section{Electroweak Baryogenesis in the MSSM}

Electroweak Baryogenesis in the MSSM is one of the most carefully
studied ideas for baryogenesis, owing to its close ties to present-day
phenomenology and accelerator searches [24--29].  Its basic
mechanism [30] is intuitively easy to understand: particles
interact in a CP-violating manner with bubble walls, which form during
the first order electroweak phase transition, when the temperature of
the universe was near $T = 100$ GeV.  This causes a buildup of a
left-handed quark density in excess of that of the corresponding
antiquarks, and an equal and opposite right-handed asymmetry, so that
there is initially no net baryon number.  The left-handed quark
asymmetry biases anomalous sphaleron interactions, present within the
standard model, to violate baryon number preferentially to create a net
quark density.  The resulting baryon asymmetry of the universe (BAU)
soon falls inside the interiors of the expanding bubbles, where the
sphaleron interactions are shut off (provided that $\langle H\rangle >
T$), and thus baryon number is safe from subsequent sphaleron-induced
relaxation to zero.

However, the correct way to treat the generation of the chiral quark
asymmetry in front of the bubble wall is controversial.  In the
simplest model, the top quark has a spatially varying complex mass,
$m(x) = |m(x)| e^{i\theta(x)}$, which gives rise to CP-violating
quantum mechanical reflection of quarks as they pass through the bubble
wall.  It also induces a CP-violating {\it classical} force on the
quarks.  In ref.\ [31] it was shown that the latter is the more
appropriate treatment when the bubble wall thickness $l_w$ is large
compared to the inverse temperature, which is the case in the MSSM: $l_w
= (10\pm 4)/T$ [29].  Furthermore it is possible to rigorously derive
the way in which the CP violating force influences the particle transport,
using the Boltzmann equation; no such complete derivation yet exists in the
quantum reflection formalism.  

In the MSSM, the top quark mass does not have a CP-violating phase; however
the charginos do; their mass matrix has the form
\beq
  \overline\psi_R M_\chi \psi_L = (\overline{\widetilde w^{^+}},\
  \overline{\widetilde h^{^+}_{2}} )_{R}
  \left(\begin{array}{cc} 
             m_2 & g H_2(x) \\
           g H_1(x) & \mu 	
        \end{array}\right)
  \left(\begin{array}{c}
         \widetilde w^{^+} \\ 
         \widetilde h^{^+}_{1} 
        \end{array}
  \right)_{\!\!L} + {\rm h.c.}
\label{eq:chmass}
\eeq
Since the SUSY parameters $\mu$ and $m_2$ can be complex, the mass eigenstates
can have spatially varying complex phases in the wall.  There thus arises
a classical force which separates the two kinds of Higgsinos, $\tilde h_{2,L}^+$
and $\tilde h_{1,L}^+$, in front of the wall.  This kind of asymmetry is not
enough to bias the sphaleron interactions, but scatterings, such as
$\tilde h_2 \tilde g \to t_L \bar t_R$, will partially convert the Higgsino
asymmetry into a chiral quark asymmetry, $n_{q_L}$.  Once the latter is determined,
it is straightforward to integrate the rate of baryon violation by sphalerons,
governed by the equation
\beq
	{d n_{\sss B}\over dt} = 27\, {\Gamma_{\rm sph}\over T^3}\, n_{q_{_L}}.
\eeq
The rate of sphaleron interactions per unit volume has been measured by
lattice simulations to be $\Gamma_{\rm sph} = (20\pm 2)\alpha_w^5 T^4$
[32].

To determine $n_{q_L}$, we must solve a set of coupled diffusion equations
for the various species $i$ of particles in the plasma.  They have the form
\beq
-D_i n_i'' - v_w n_i + \Gamma_{ijk}(n_i-n_j-n_k)  = S_i,
\eeq
where $D_i$ is a diffusion coefficient (of order the inverse mean free path), 
$v_w$ is the bubble wall
velocity, $\Gamma_{ijk}$ is the rate of interactions of the type $i\to
j+k$ (given as an example), and $S_i$ is a CP-violating source term
arising from the force on the particles or from quantum reflections.
It is possible to show that the source term in the Higgsino diffusion
equation is related to the force $F$ by the thermal averages
\beq
S(x) = - {v_w D \over \avg{\vec v^{\,2}}} 
\avg{v_x F(x)}',
\label{eq:source}
\eeq
and, by solving the Dirac equation in the WKB approximation, 
the CP-violating part of the force is related to the complex 
Higgsino mass $m e^{i\theta}$ by
$F = {(s/ 2 E^2)}({m^2\theta'})',$
where $s=\pm 1$ is the spin.  Moreover the combination $m^2\theta'$ is
given by
\beq
  m^2\theta' =  
        {g^2{\rm Im}(m_2\mu)\over 2(m^2_+-m^2_-)}
        \left( H_1 H_2' + H_1' H_2 \right),
\label{eq:mssm_angle}
\eeq
where $m^2_\pm$ are the two eigenstates of the mass matrix in
(\ref{eq:chmass}).  A remarkable feature of this expression is the
relative $+$ sign between $H_1 H_2'$ and $H_1' H_2$.  Other authors
[24,26] had previously obtained a source proportional to $H_1
H_2'$ {\it minus} $H_1' H_2$, which is highly suppressed, because
$H_1/H_2$ tends to be constant within the wall, to within a part in
$10^2$ or $10^3$ [33,10]. The origin of the discrepancy is that
(for technical reasons) the previous authors considered only the linear
combination of Higgsino densities $n_{\tilde h_1} - n_{\tilde h_2}$,
whereas our classical force is providing a source for $n_{\tilde h_1} +
n_{\tilde h_2}$ [29].

By solving the difffusion equations and numerically integrating the
baryon violation rate equation, we obtained the baryon asymmetry as a
function of the model-dependent parameters of the MSSM [29].
Figure \ref{fig3}(a) shows how $\eta_{10}$ varies with the bubble wall
velocity $v_w$ in a typical case, for the allowed range of values of
the bubble wall thickness, where we assumed maximal CP violation,
Im$(m_2\mu) = |m_2\mu|$.  Since typically $\eta_{10}$ is $\sim 1000$,
but we only need $(2-3)$, we see that the CP-violating phase need not
be maximal, but could be $(2-3)\times 10^{-3}$.  This is good news, since
the neutron and electric dipole moment searches give constraints which
are of this order, unless some kind of fine tuning is invoked.  Baryon
production peaks at small wall velocities near $v_w = 10^{-2}$.
Interestingly, recent estimates of $v_w$ in the MSSM give values which
are this small [34].  Figure \ref{fig3}(b) shows the contours of
constant CP phase which yield $\eta_{10} =3$ in the chargino mass
parameter plane.

\begin{figure}
\centerline{\epsfxsize=2.45in\epsfbox{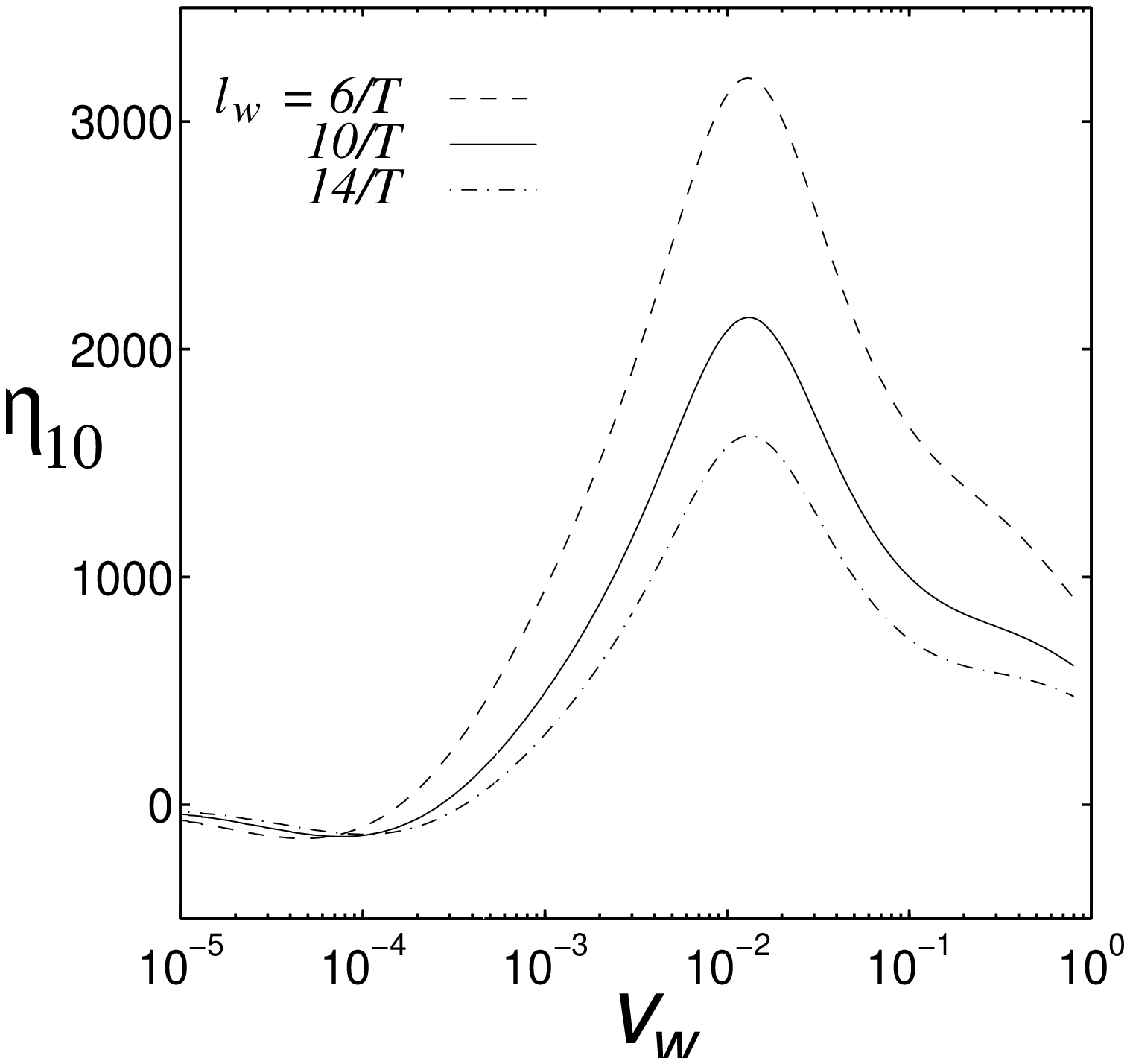}\hfill
\epsfxsize=2.4in\epsfbox{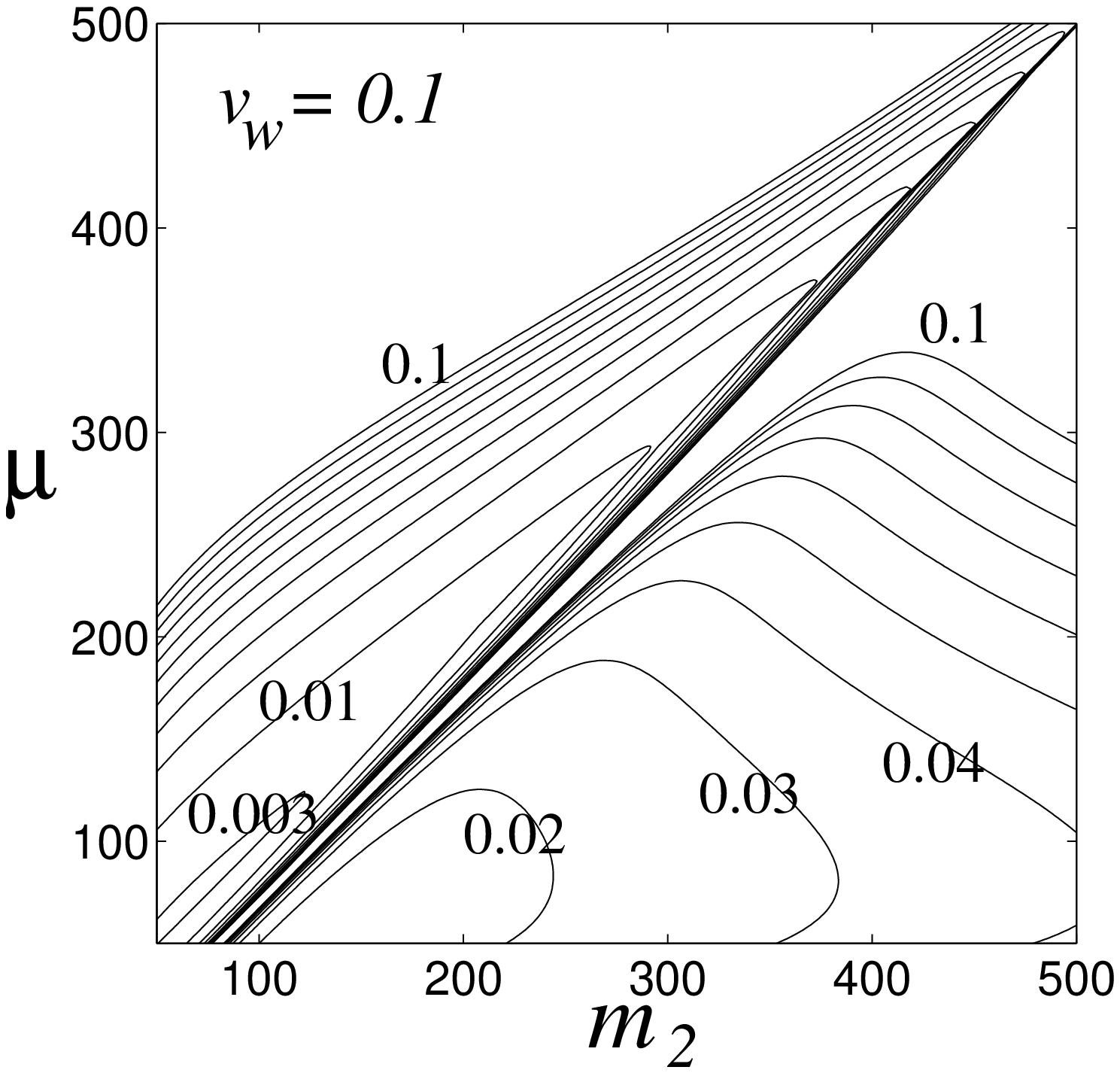}}
\vspace{0.2in}
\caption{(a) Baryon asymmetry versus wall velocity for maximal
CP violation and three different wall thicknesses; and (b) contours
of constant CP-violating phase giving the desired baryon asymmetry,
in the chargino mass parameter plane.}
\label{fig3}
\end{figure}

\section{Conclusions}

In this talk I have discussed one of the main ``applications'' of
cosmological phase transitions, baryogenesis.  Although there are many
imaginative ideas for getting the baryon asymmetry, the most mainstream
ones are those that are most closely related to phenomenology.
Baryogenesis from leptogenesis is appealing because it can potentially
make contact with neutrino masses.  Electroweak baryogenesis is
indirectly testable by searches for Higgs bosons, the top squark,
charginos, neutralinos and electric dipole moments.  We have seen that
it is relatively easy to generate a large enough baryon asymmetry in
the electroweak model with the MSSM, but it is not so easy to get a
strong enough phase transition to safeguard it against washout by
sphalerons: a large hierarchy between the left and right stop masses is
needed.  It would be nice to have some more robust way of strengthening the
transition.  One possibility is adding a singlet Higgs field [35].
An interesting proposal is to reheat after inflation to temperatures below
the electroweak transition, but rely upon nonequilibrium effects, similar
to parametric resonance, to produce sphalerons [36].
Another idea is to modify the expansion rate of the universe instead.
It has been noted that extra dimensions could have this effect at sufficiently
early times [37].

\end{document}